\documentclass[conference]{IEEEtran}

\usepackage[T1]{fontenc}
\usepackage{amsmath}
\usepackage{amssymb}
\usepackage{stmaryrd}
\usepackage{epic,eepic}
\usepackage{theorem}
\usepackage{pifont}  %needed by dingautolist
\usepackage{euscript}
\usepackage{calc}
\usepackage[titles]{tocloft}
\usepackage[usenames]{color}          % Need the color package
\usepackage{xcolor}
\definecolor{Brown}{rgb}{0.55,0.0,0.10}
\definecolor{dgreen}{rgb}{0.00,0.56,0.00}
\definecolor{vertmoinsfonce}{rgb}{0.00,0.50,0.00}
\definecolor{vert}{rgb}{0.00,0.60,0.00}
\definecolor{llightggray}{rgb}{0.97,0.97,0.97}
\definecolor{lightggray}{rgb}{0.9,0.9,0.9}
\definecolor{ggray}{rgb}{0.5,0.5,0.5}
\definecolor{darkggray}{rgb}{0.25,0.25,0.25}
\definecolor{ddarkggray}{rgb}{0.1,0.1,0.1}
\definecolor{bleu}{rgb}{0.00,0.00,1.00}

\usepackage{epsf}           % dessin
\usepackage{graphicx}       % dessin
\usepackage{epsfig}         % dessin
\usepackage{pstricks}
\usepackage{pstricks-add}
\usepackage{pst-plot}
\usepackage{dsfont}
\theoremheaderfont{\color{black}\normalfont\bfseries}

\newtheorem{theorem}{Theorem}[section]
\newtheorem{definition}[theorem]{Definition}

\theoremstyle{plain}{\theorembodyfont{\rmfamily}%
}
\theoremstyle{plain}{\theorembodyfont{\rmfamily}%
}
\theoremstyle{plain}{
\theorembodyfont{\rmfamily}

	}

\newcommand{\R}{\mathbb{R}}

\newcommand{\N}{\mathbb{N}}

\newcommand{\E}{\mathbb{E}}
\newcommand{\Q}{\mathbb{Q}}

\newcommand{\C}{\mathcal{C}}

\newcommand{\QQ}{\mathcal{Q}}

\newcommand{\D}{\mathcal{D}}

\newcommand{\PP}{\mathcal{P}}

\newcommand{\mc}{\mathcal}

{\Large\normalsize}
{\Large\normalsize}

\ifCLASSINFOpdf
\else
\fi
\begin{document}

%\title{Empirical Coordination for an Information Source Correlated with the Channel State}
%\title{Correlation between Channel State and Information Source for Empirical Coordination}
%\title{Channel State Correlated with Information Source under Empirical Coordination Constraint}
\title{Empirical Coordination with Channel Feedback and Strictly Causal or Causal Encoding}


\author{\IEEEauthorblockN{Ma\"{e}l Le Treust}\\
\IEEEauthorblockA{
ETIS, UMR 8051 / ENSEA, Université Cergy-Pontoise, CNRS,\\% F-95000, Cergy
%ETIS, CNRS UMR8051, ENSEA, Université de Cergy-Pontoise,\\
6, avenue du Ponceau,\\
95014 CERGY-PONTOISE CEDEX,\\
FRANCE\\
Email: mael.le-treust@ensea.fr}
}

%\author{Maël~Le~Treust,~\IEEEmembership{Member,~IEEE,}
%        Abdellatif Zaidi,~\IEEEmembership{Member,~IEEE},
%        \thanks{Laboratoire d'Informatique de l'Institut Gaspard Monge,
%Universit\'{e} Paris-Est Marne La Vall\'{e}e,77454, Marne La Vall\'{e}e Cedex 2, France.
%Email: \{mael.letreust\},\{abdellatif.zaidi\}@univ-mlv.fr}}

%% The paper headers
%\markboth{Journal of \LaTeX\ Class Files,~Vol.~6, No.~1, January~2007}%
%{Shell \MakeLowercase{\textit{et al.}}: Bare Demo of IEEEtran.cls for Journals}

\maketitle

\begin{abstract}

In multi-terminal networks, feedback increases the capacity region and helps communication devices to coordinate.  In this article, we deepen the relationship between coordination and feedback by considering a point-to-point scenario with an information source and a noisy channel. Empirical coordination is achievable if the encoder and the decoder can implement  sequences of symbols that are jointly typical for a target probability distribution. We investigate the impact of feedback when the encoder has strictly causal or causal observation of the source symbols. For both cases, we characterize the optimal information constraints and we show that feedback improves coordination possibilities. Surprisingly, feedback also reduces the number of auxiliary random variables and simplifies the information constraints. For empirical coordination with strictly causal encoding and feedback, the information constraint does not involve auxiliary random variable anymore.

%%%%

\end{abstract}


% For peer review articles, you can put extra information on the cover
% \fi
%
% For peerreview articles, this IEEEtran command inserts a page break and
\IEEEpeerreviewmaketitle

\section{Introduction}\label{sec:Introduction}

%%%%%%%%%%%%%%%%%%%%%%%%%%%

Feedback does not increase the capacity of a memoryless channel \cite{Shannon-zero-error56}. However, it has a significant impact when considering problems of empirical coordination.
% \cite{KramerSavari07}, \cite{CuffPermuterCover10},  \cite{Cuff(ImplicitCoordination)10} for a point-to-point scenario with a noisy channel. 
 In this framework, encoder and decoder are considered as autonomous agents  \cite{GoHerNey06}, that implement a coding scheme in order to coordinate their sequences of  actions, \textit{i.e.} channel inputs and decoder outputs, with a sequence of source symbols.
 %, also called actions \cite{GoHerNey06}. 
 The problem of empirical coordination \cite{KramerSavari07}, \cite{CuffPermuterCover10},  \cite{Cuff(ImplicitCoordination)10} consists in determining the set of joint probability distributions, that are achievable for empirical frequencies of symbols. 
Empirical coordination provides a single-letter solution that simplifies the analysis of optimization problems such as minimal source distortion, minimal channel cost or maximal utility function of a decentralized communication network \cite{LeTreust(EmpiricalCoordination)14}. For example, the optimal distortion level is the minimum of the expected distortion function, taken over the set of achievable joint probability distributions.

% XXX
% The objective is to provide a single-letter formulation for optimization problems such as minimal source distortion, minimal channel cost or maximal utility function of decentralized communication network \cite{LeTreust(EmpiricalCoordination)14}. It remains to optimize the expectation of the criteria (distortion/cost/utility) over the set of achievable joint probability distributions, in order to characterize the optimal solution.

In the framework of multi-terminal networks, feedback increases the capacity region of the multiple-access channel \cite{GaarderWolf(FeddbackMAC)75}, \cite{Ozarow(Feedback)84} and of the broadcast channel \cite{Dueck(Feedback)79}, \cite{OzarowLeung(FeedbackBC)84}. In the literature of game theory, feedback is considered from a strategic point-of-view. In \cite{GoHerNey06}, a player observes the past actions of another player through a monitoring structure involving perfect or imperfect feedback. In \cite{LetreustZaidiLasaulce(Allerton)11}, the authors investigate a four-player coordination game with imperfect feedback and provide a subset of achievable joint probability distributions. Empirical coordination is a first step toward a better understanding of decentralized communication network. The set of achievable joint distributions was characterized for strictly causal and causal decoding in \cite{LeTreust(EmpiricalCoordination)14}, with two-sided state information in \cite{LeTreust(ISIT-TwoSided)15} and with feedback from the source in \cite{LarrousseLasaulceBloch(IT)14}.  From a practical perspective, coordination with polar codes was considered in \cite{BlochLuzziKliewer12}. Lossless decoding with correlated information source and channel states is solved in \cite{LeTreust(CorrelationITW)14}. Empirical coordination for multi-terminal source coding is treated in \cite{BereyhiBahramiMirmohseniAref13} and in \cite{GoldfeldPermuterKramer(ISIT)14}.

% the authors characterize the set of achievable joint distributions by considering strictly causal and causal decoding without feedback and with feedback from the source. Empirical coordination with strictly causal decoding is characterized in without feedback and in 

\begin{figure}[!ht]
\begin{center}
\psset{xunit=0.9cm,yunit=0.9cm}
\begin{pspicture}(0,-0.5)(8.5,1)
\pscircle(0,0.5){0.45}
\psframe(2,0)(3,1)
\pscircle(5,0.5){0.45}
\psframe(7,0)(8,1)
\psline[linewidth=1pt]{->}(0.5,0.5)(2,0.5)
\psline[linewidth=1pt]{->}(3,0.5)(4.5,0.5)
\psline[linewidth=1pt]{->}(5.5,0.5)(7,0.5)
\psline[linewidth=1pt]{->}(8,0.5)(9,0.5)
%\psline[linewidth=1pt]{->}(0,0)(0,-0.5)(5,-0.5)(5,0)
%\psline[linewidth=1pt]{->}(2.5,-0.5)(2.5,0)
\psline[linewidth=1pt]{->}(5,0)(5,-0.5)(2.5,-0.5)(2.5,0)
\rput[u](1,0.8){$U^{i-1}$}
\rput[u](3.75,0.8){$X_i$}
\rput[u](6.25,0.8){$Y^{n}$}
\rput[u](8.5,0.8){$V^n$}
\rput[u](3.75,-0.2){$Y^{i-1}$}
\rput(0,0.5){$\PP_{\sf{u}}$}
\rput(2.5,0.5){$\C$}
\rput(5,0.5){$\mc{T}$}
\rput(7.5,0.5){$\D$}
\end{pspicture}
\caption{Strictly causal encoding function with feedback $f_i: \mc{U}^{i-1} \times \mc{Y}^{i-1} \rightarrow \mc{X}$, for all $i\in\{1,\ldots,n\}$ and non-causal decoding function $g: \mc{Y}^n \rightarrow \mc{V}^n$ .}\label{fig:StrictlyCausalEncFeedbacks}
\end{center}
\end{figure}
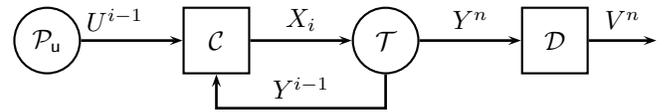

\vspace{-0.3cm}

In this article, we consider the point-to-point scenario of \cite{CuffSchieler11} with channel feedback, as represented by Fig. \ref{fig:StrictlyCausalEncFeedbacks} and \ref{fig:CausalEncFeedbacks}. The encoder has perfect feedback from the channel and strictly causal or causal observation of the symbols of source. In both cases, we characterize the set of achievable joint probability distributions over the symbols of source and channel. We show that the information constraints are larger than the ones stated in \cite{CuffSchieler11}. Surprisingly, feedback also reduces the number of auxiliary random variables and simplifies the information constraints. For empirical coordination with strictly causal encoding and feedback, the information constraint does not involve auxiliary random variable anymore. There is an analogy with strictly causal decoding \cite{LeTreust(EmpiricalCoordination)14}, %\cite{SamsonBenjaminISIT13}, 
\cite{LarrousseLasaulceBloch(IT)14}, since no auxiliary random variable is needed  when the decoder has feedback from the source.
%also only involves the random variables of the source and the channel. 
Feedback allows to remove auxiliary random variables of information constraints, for empirical coordination problems.

System model and definitions are stated in Sec. \ref{sec:ModelDefinition} and characterizations of achievable joint distributions are stated in Sec. \ref{sec:MainResults}. Comparison with previous works and an example are  stated in Sec. \ref{sec:ComparisonPreviousResults}  and \ref{sec:NumericalExample}. 
%Sec. \ref{sec:ComparisonPreviousResults} presents  a comparison between the main results with previous results without feedback. An example of binary source and binary symmetric channel is provided in Sec. 
Conclusions and sketches of proofs are stated in Sec. \ref{sec:Conclusion} and in Appendix  \ref{sec:ProofFeedbacks2}, \ref{sec:ProofFeedbacks3}, \ref{sec:ProofFeedbacks4}.

\section{System model}\label{sec:ModelDefinition}

%\small
Figure \ref{fig:StrictlyCausalEncFeedbacks} represents the problem under investigation.
Random variable $U$ is denoted by capital letter, lowercase letter $u\in\mc{U}$ designates the realization and $\mc{U}^n$ corresponds to the $n$-time cartesian product. $U^n$, $X^n$, $Y^n$,  $V^n$ stands for sequences of random variables of source symbols $u^n=(u_1,\ldots,u_n)\in\mc{U}^n$, inputs of the channel  $x^n\in\mc{X}^n$, outputs of the channel $y^n\in\mc{Y}^n$ and decoder's output $v^n\in\mc{V}^n$. The sets $\mc{U}$, $\mc{X}$, $\mc{Y}$, $\mc{V}$ are discrete. The set of  probability distributions $\PP(X)$ over $\mc{X}$ is denoted by $\Delta(\mc{X})$. Notation $||\QQ - \PP||_{\sf{tv}}= 1/2\cdot \sum_{x\in\mc{X}} |\QQ(x) - \PP(x)|$ stands for the total variation distance between probability distributions $\QQ$ and $\PP$. %Denote by $\UN(y|x)$, the indicator function equal to 1 if $y = x$ and 0 otherwise. 
Notation $Y  -\!\!\!\!\minuso\!\!\!\!-X    -\!\!\!\!\minuso\!\!\!\!-  U$ stands for the Markov chain property corresponding to $\PP(y|x,u) = \PP(y|x)$ for all $(u,x,y)$. Information source is i.i.d. distributed with $\PP_{\sf{u}}$ and the channel is memoryless with transition probability $\mc{T}_{\sf{y|x}}$. Encoder $\C$ and decoder $\D$ know the statistics $\PP_{\sf{u}}$ and $\mc{T}_{\sf{y|x}}$ of the source and channel. The coding process is deterministic.

%\normalsize

%Coding Process: A sequence of source symbols $u^n\in\mc{U}^n$, of channel states $s^n\in\mc{S}^n$ and of side information at decoder $z^n\in\mc{Z}^n$ are drawn from the i.i.d. probability distribution defined by equation \eqref{eq:SourceProba}. Encoder $\C$ observes $u^n\in\mc{U}^n$, $s^n\in\mc{S}^n$ non-causally and sends a sequence of channel inputs $x^n\in\mc{X}^n$. The sequence of channel outputs $y^n \in \mc{Y}^n$ is drawn according to the discrete and memoryless transition probability defined by equation (\ref{eq:TransitionProba}).
%\begin{eqnarray}
%\mc{P}_{\sf{usz}}^{\otimes n}(u^n,s^n,z^n)= \prod_{i=1}^n \mc{P}_{\sf{usz}}(u_i,s_i,z_i),\label{eq:SourceProba}\\
%\mc{T}^{\otimes n}(y^n|x^n,s^n)= \prod_{i=1}^n \mc{T}(y_i|x_i,s_i).\label{eq:TransitionProba}
%\end{eqnarray}
%The decoder observes the sequence of channel outputs $y^n\in\mc{Y}^n$ and the sequence of side information $z^n\in\mc{Z}^n$ and returns a sequence $v^n\in\mc{V}^n$. 

\begin{definition}\label{def:Code}
A  code $c\in\mc{C}(n)$ with strictly-causal encoder and feedback is a tuple of functions $c=(\{f_i\}_{i=1}^n, g)$ defined by equations \eqref{eq:1CausalCodeSource1} and \eqref{eq:1CausalCodeSource2}:
\begin{eqnarray}
&f_i& : \mc{U}^{i-1} \times  \mc{Y}^{i-1} \longrightarrow \mc{X},\qquad i = 1,\ldots,n ,\label{eq:1CausalCodeSource1}\\
&g& : \mc{Y}^n  \longrightarrow \mc{V}^n.\label{eq:1CausalCodeSource2}
\end{eqnarray}
The  number of occurrence of symbol $u \in \mc{U}$ in sequence $u^n$ is denoted by $\textsf{N}(u|u^n)$. The empirical distribution ${Q}^n \in \Delta(\mc{U} \times \mc{X}\times \mc{Y} \times\mc{V})$ of sequences $(u^n,x^n,y^n,v^n)$
is defined by:
%\begin{small}
\begin{eqnarray}
{Q}^n(u,x,y,v) &=& \frac{\textsf{N}(u,x,y,v |u^n,x^n,y^n,v^n)}{n},  \nonumber \\
 \forall  (u,x,y,v)& \in& \mc{U}\times   \mc{X} \times \mc{Y} \times\mc{V}. \label{eq:EmpiricalDistribution}
\end{eqnarray}
%\end{small}
Fix a target probability distribution $\QQ \in \Delta(\mc{U} \times \mc{X} \times \mc{Y} \times \mc{V}  )$, the error probability of the code $c\in\mc{C}(n)$ is defined by:
\begin{eqnarray}
\PP_{\textsf{e}}(c) = \PP_c\bigg(\Big|\Big|Q^n - \QQ \Big|\Big|_{\sf{tv}}\geq \varepsilon\bigg),\label{eq:ErrorProba}
\end{eqnarray}
where $Q^n \in \Delta(\mc{U} \times \mc{X}\times \mc{Y} \times\mc{V})$ is the random variable of the empirical distribution induced by the probability distributions $\PP_{\sf{u}}$, $\mc{T}_{\sf{y|x}}$ and the code $c \in \mc{C}(n)$.
\end{definition}

\begin{definition}
The probability distribution $\QQ \in  \Delta(\mc{U} \times  \mc{X} \times \mc{Y} \times \mc{V}  )$ is achievable if for all $\varepsilon>0$, there exists a $\bar{n}\in \N$ s.t. for all $n \geq \bar{n}$, there exists a code $c\in\mc{C}(n)$ that satisfies:
\begin{eqnarray}
\PP_{\textsf{e}}(c)  = \PP_c\bigg(\Big|\Big|Q^n - \QQ \Big|\Big|_{\sf{tv}}\geq \varepsilon\bigg) \leq  \varepsilon.
\end{eqnarray}
\end{definition}
The error probability $\PP_{\textsf{e}}(c)$ is small if the total variation distance between the empirical frequency of symbols $Q^n(u,x,y,v)$ and the target probability distribution $\QQ(u,x,y,v)$ is small, with large probability. In that case, the sequences of symbols $(U^n, X^n,Y^n,V^n)\in A_{\varepsilon}^{{\star}{n}}(\QQ)$ are jointly typical, \textit{i.e.} coordinated, for the target probability distribution $\QQ$ with large probability. %This means that the sequences of symbols are coordinated empirically for probability distribution $\QQ$. 

As mentioned in \cite{LeTreust(EmpiricalCoordination)14} and \cite{LeTreust(CorrelationITW)14}, the performance of the coordination can be evaluated using an objective function  $\Phi : \mc{U} \times  \mc{X}  \times \mc{Y} \times \mc{V} \mapsto \R$. We denote by $\mc{A}^{\star}$, the set of joint probability distributions $\QQ \in \mc{A}^{\star} $ that are achievable.  Based on the expectation $ \E_{\QQ \in \mc{A}^{\star} }\Big[\Phi(U,X,Y,V) \Big]$, it is possible to derive the minimal channel cost $\Phi(u,x,y,v) = c(x)$, the minimal distortion level $\Phi(u,x,y,v) = d(u,v)$ or the maximal utility of a decentralized network \cite{GoHerNey06}, using a single-letter characterization.

%%%%%%%%%%%%%%%%%%%%%%%%%%%%%%%%%%%%%%
%%%%%%%%%%%%%%%%%%%%%%%%%%%%%%%%%%%%%%

%\newpage

\section{Characterization of achievable distributions}\label{sec:MainResults}

This section presents the two main results of this article. Theorem \ref{theo:feedbacks} characterizes of the set of achievable joint probability distributions for strictly causal encoding with feedback, represented in Fig. \ref{fig:StrictlyCausalEncFeedbacks}.

\begin{theorem}[Strictly causal encoding with feedback] 
$\qquad\qquad\qquad\qquad\qquad\qquad\qquad\qquad\qquad\qquad\qquad\qquad\qquad\qquad\qquad$\label{theo:feedbacks}
$1)$ If the joint probability distribution $\QQ(u,x,y,v)$ is achievable, then it decomposes as follows:
\begin{eqnarray}
\begin{cases}
\QQ(u) = \PP_{\sf{u}}(u), \;\;& \QQ(y|x) = \mc{T}(y | x),\\
U \text{ independent of } X,\;\;\ &Y -\!\!\!\!\minuso\!\!\!\!- X    -\!\!\!\!\minuso\!\!\!\!-   U.
\end{cases}
%\begin{cases}
%&\QQ(u) = \PP_{\sf{u}}(u),\\
%& \QQ(y|x) = \mc{T}(y | x),\\
%&U \text{ independent of } X,\\
%&Y -\!\!\!\!\minuso\!\!\!\!- X    -\!\!\!\!\minuso\!\!\!\!-   U ,
%%&\PP_{\sf{usz}}(u,s,z)   \otimes   \QQ(x|u,s)  \otimes \mc{T}(y | x ,s)  \otimes \QQ(v|u,s,z,x,y)\\
%%& \qquad\qquad\qquad\qquad\qquad \qquad\qquad\qquad\quad \; \text{ is achievable}.\nonumber
%\end{cases}
\end{eqnarray}
%and $\PP_{\sf{u}}(u)   \otimes   \QQ(x)  \otimes \mc{T}(y | x )  \otimes \QQ(v|u,x,y)$ is achievable.\\
2) Joint probability distribution $\PP_{\sf{u}}(u)   \otimes   \QQ(x)  \otimes \mc{T}(y | x )  \otimes \QQ(v|u,x,y)$ is achievable if:
\begin{eqnarray}
 I(X ; Y  )   -  I( U;V  | X , Y )   > 0 ,  \label{eq:Feedbacks1}
\end{eqnarray}
3) Joint probability distribution $\PP_{\sf{u}}(u)   \otimes   \QQ(x)  \otimes \mc{T}(y | x )  \otimes \QQ(v|u,x,y) $ is not achievable if:
\begin{eqnarray}
 I(X ; Y  )   -  I( U;V  | X , Y )   < 0 ,  \label{eq:Feedbacks2}
\end{eqnarray}
\end{theorem}
Sketch of proof of Theorem \ref{theo:feedbacks} is stated in Appendix \ref{sec:ProofFeedbacks2}. % and the full version is available in \cite{LeTreust(RapportISITfeedbacks)15}. 
%No auxiliary random variable is involved in the information constraint \eqref{eq:Feedbacks1}. 
%The auxiliary random variable of Theorem 3 in \cite{CuffSchieler11} is replaced by $V$ and the 
Equation \eqref{eq:Feedbacks1} comes from Theorem 3 in \cite{CuffSchieler11} by replacing the auxiliary random variable by decoder's output $V$ and the observation of the encoder by the pair  of information source and channel feedback $(U,Y)$.

\begin{figure}[!ht]
\begin{center}
\psset{xunit=0.9cm,yunit=0.9cm}
\begin{pspicture}(0,-0.3)(8.5,1)
\pscircle(0,0.5){0.45}
\psframe(2,0)(3,1)
\pscircle(5,0.5){0.45}
\psframe(7,0)(8,1)
\psline[linewidth=1pt]{->}(0.5,0.5)(2,0.5)
\psline[linewidth=1pt]{->}(3,0.5)(4.5,0.5)
\psline[linewidth=1pt]{->}(5.5,0.5)(7,0.5)
\psline[linewidth=1pt]{->}(8,0.5)(9,0.5)
%\psline[linewidth=1pt]{->}(0,0)(0,-0.5)(5,-0.5)(5,0)
%\psline[linewidth=1pt]{->}(2.5,-0.5)(2.5,0)
\psline[linewidth=1pt]{->}(5,0)(5,-0.5)(2.5,-0.5)(2.5,0)
\rput[u](1,0.8){$U^{i}$}
\rput[u](3.75,0.8){$X_i$}
\rput[u](6.25,0.8){$Y^{n}$}
\rput[u](8.5,0.8){$V^n$}
\rput[u](3.75,-0.2){$Y^{i-1}$}
\rput(0,0.5){$\PP_{\sf{u}}$}
\rput(2.5,0.5){$\C$}
\rput(5,0.5){$\mc{T}$}
\rput(7.5,0.5){$\D$}
\end{pspicture}
\caption{Causal encoding function with feedback $f_i: \mc{U}^{i} \times \mc{Y}^{i-1} \rightarrow \mc{X}$, for all $i\in\{1,\ldots,n\}$ and non-causal decoding function $g: \mc{Y}^n \rightarrow \mc{V}^n$ .}\label{fig:CausalEncFeedbacks}
\end{center}
\end{figure}
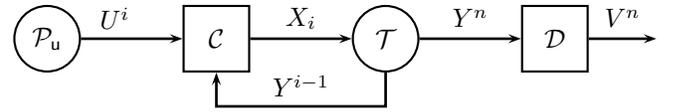

A causal encoding function is defined by $f_i : \mc{U}^{i} \times  \mc{Y}^{i-1} \rightarrow \mc{X}, \; \forall i\in \{1,\ldots, n\}$. Theorem \ref{theo:Causalfeedbacks} characterizes of the set of achievable joint probability distributions for causal encoding with feedback, represented in Fig. \ref{fig:CausalEncFeedbacks}.

\vspace{-0.5cm}

\begin{theorem}[Causal Encoding with Feedback] 
$\qquad\qquad\qquad\qquad\qquad\qquad\qquad\qquad\qquad\qquad\qquad\qquad\qquad\qquad\qquad$\label{theo:Causalfeedbacks}
$1)$ If the joint probability distribution $\QQ(u,x,y,v)$ is achievable, then it decomposes as follows:
\begin{eqnarray}
\QQ(u) = \PP_{\sf{u}}(u), \;\;  \QQ(y|x) = \mc{T}(y | x),\;\; Y -\!\!\!\!\minuso\!\!\!\!- X    -\!\!\!\!\minuso\!\!\!\!-   U,
%\begin{cases}
%&\QQ(u) = \PP_{\sf{u}}(u),\\
%& \QQ(y|x) = \mc{T}(y | x),\\
%%&U \text{ independent of } X,\\
%&Y -\!\!\!\!\minuso\!\!\!\!- X    -\!\!\!\!\minuso\!\!\!\!-   U ,
%%&\PP_{\sf{usz}}(u,s,z)   \otimes   \QQ(x|u,s)  \otimes \mc{T}(y | x ,s)  \otimes \QQ(v|u,s,z,x,y)\\
%%& \qquad\qquad\qquad\qquad\qquad \qquad\qquad\qquad\quad \; \text{ is achievable}.\nonumber
%\end{cases}
\end{eqnarray}
%and $\PP_{\sf{u}}(u)   \otimes   \QQ(x|u)  \otimes \mc{T}(y | x )  \otimes \QQ(v|u,x,y)$ is achievable.\\
2) Joint probability distribution $\PP_{\sf{u}}(u)   \otimes   \QQ(x|u)  \otimes \mc{T}(y | x )  \otimes \QQ(v|u,x,y)$ is achievable if:
\begin{eqnarray}
\max_{{\QQ}\in \Q} \bigg( I(W ; Y  )   -  I( U;V  | W , Y )  \bigg)   > 0 ,  \label{eq:FeedbacksCausal1}
\end{eqnarray}
3) Joint probability distribution $\PP_{\sf{u}}(u)   \otimes   \QQ(x|u)  \otimes \mc{T}(y | x )  \otimes \QQ(v|u,x,y) $ is not achievable if:
\begin{eqnarray}
\max_{{\QQ}\in \Q} \bigg( I(W ; Y  )   -  I( U;V  | W , Y ) \bigg)   < 0 ,  \label{eq:FeedbacksCausal2}
\end{eqnarray}
where $\Q$ is the set of probability distributions ${\QQ}   \in  \Delta(\mc{U} \times \mc{W} \times \mc{X} \times \mc{Y}\times \mc{V}  )$ with auxiliary random variable $W$ that satisfies:
\begin{small}
\begin{eqnarray*}
\begin{cases}
\sum_{w\in \mc{W}} {\QQ}(u,w,x,y,v) \\
\quad =  \PP_{\sf{u}}(u)   \otimes   \QQ(x|u)\otimes \mc{T}(y | x)   \otimes   \QQ(v|u,x,y) , \\
U \text{ independent of } W,\\
Y -\!\!\!\!\minuso\!\!\!\!- X -\!\!\!\!\minuso\!\!\!\!-  (U,W) ,\\
V -\!\!\!\!\minuso\!\!\!\!- (U,Y,W ) -\!\!\!\!\minuso\!\!\!\!- X.
\end{cases}
\end{eqnarray*}
\end{small}
The probability distribution $\QQ \in \Q$ decomposes as follows:
\begin{eqnarray*}
 \PP_{\sf{u}}(u)   \otimes   \QQ(w) \otimes   \QQ(x|u,w) \otimes      \mc{T}(y | x )   \otimes   \QQ(v|u,y,w).
 \end{eqnarray*}
The support of $W$ is bounded by $|\mc{W}| \leq   |\mc{U}\times \mc{X}\times \mc{Y} \times \mc{V} |  +2$.
\end{theorem}
Sketch of proofs of Theorem \ref{theo:Causalfeedbacks} are stated in Appendix \ref{sec:ProofFeedbacks3} and \ref{sec:ProofFeedbacks4}. % and the full version is available in \cite{LeTreust(RapportISITfeedbacks)15}.
The random variable $V$ is directly correlated with the pair $(U,Y)$ of source and channel output. Feedback implies that $V$ is extracted from the Markov chain  $Y -\!\!\!\!\minuso\!\!\!\!- X -\!\!\!\!\minuso\!\!\!\!-  (U,W)$ of the memoryless channel.

%%%%%%%%%%%%%%%%%%%%%%%%%%%%%%%%%%%%%%
%%%%%%%%%%%%%%%%%%%%%%%%%%%%%%%%%%%%%%

\section{feedback improves empirical coordination}\label{sec:ComparisonPreviousResults}

In this section, we investigate the impact of the feedback on the set of achievable joint distributions stated in Theorems \ref{theo:feedbacks} and \ref{theo:Causalfeedbacks}. Considering strictly causal encoding, we evaluate the difference between information constraint stated in equation \eqref{eq:Feedbacks1} and the one stated in Theorem 3 in \cite{CuffSchieler11} without feedback.
%compare the information constraints of Theorems \ref{theo:feedbacks} and \ref{theo:Causalfeedbacks} to the ones stated in \cite{CuffSchieler11} for strictly causal and causal encoding without feedback. 
%Regarding the case of strictly causal encoding, the difference between information constraint stated in equation \eqref{eq:Feedbacks1} and the one stated in Theorem 3 in \cite{CuffSchieler11} is equal to:
\begin{eqnarray}
 &&I(X ; Y  )   -  I( U;V  | X , Y )  \label{eq:SCEFeedbacks} \\
 &-& \max_{{\QQ}\in \Q_{\sf{se}} } \bigg( I(X ; Y  )   -  I( U  ; W_2 | X )   \bigg)\label{eq:SCE} \\
 &=&\min_{{\QQ}\in \Q_{\sf{se}} }  I( U  ; W_2 | X )   - I( U;V  | X , Y )  \\
  &=& H( U  | V,X , Y )  - \max_{{\QQ}\in \Q_{\sf{se}} }  H(U | X , W_2)\geq 0 .\label{eq:comparison1}
\end{eqnarray}
$\Q_{\sf{se}}$ is the set of probability distributions ${\QQ}   \in  \Delta(\mc{U}     \times \mc{W}_2 \times \mc{X} \times \mc{Y}\times \mc{V}  )$ with auxiliary random variable $W_2$ that satisfies:
\begin{eqnarray*}
\PP_{\sf{u}}(u)   \otimes   \QQ(x) \otimes   \QQ(w_2|u,x) \otimes   \mc{T}(y | x )   \otimes   \QQ(v|y,x,w_2).
 \end{eqnarray*}
$\bullet$ Equation \eqref{eq:comparison1} is equal to zero if $(U,V)$ is independent of $(X,Y)$, this corresponds to the lossy transmission without coordination in which the feedback does not increase the channel capacity \cite{Shannon-zero-error56}. \\
$\bullet$ Equation \eqref{eq:comparison1} is equal to zero when the decoder output $V$ is empirically coordinated with $(U,X)$ and not with the channel output $Y$, because in that case $W_2=V$. %By replacing $U$ by $(U,Y)$ and $W_2$  by $V$ in equation \eqref{eq:SCE}, this gives the information constraint \eqref{eq:SCEFeedbacks} with the feedback. 
Since the auxiliary random variable $W_2$ should satisfy $\QQ(v|y,x,u) = \sum_{w_2 \in \mc{W}_2}  \QQ(w_2|u,x) \cdot \QQ(v|y,x,w_2)$, %no auxiliary random variable is involved, 
equation \eqref{eq:SCEFeedbacks} provides an upper bound to equation \eqref{eq:SCE} that is easier to evaluate
 
%There is a strong analogy between strictly causal encoding with/without channel feedback in equation  \eqref{eq:SCEFeedbacks} and \eqref{eq:SCE} and strictly causal decoding with/without feedback from the information source in equation \eqref{eq:SCDFeedbacks} and \eqref{eq:SCD}, characterized in \cite{SamsonBenjaminISIT13}, \cite{LarrousseLasaulceBloch(IT)14} and \cite{LeTreust(EmpiricalCoordination)14}. 

There is a strong analogy between strictly causal \textit{encoding} with channel feedback and strictly causal \textit{decoding} with source feedback.
%, characterized in \cite{LeTreust(EmpiricalCoordination)14}, \cite{SamsonBenjaminISIT13} and \cite{LarrousseLasaulceBloch(IT)14}. 
Equation \eqref{eq:SCD} corresponds to strictly causal decoding without feedback from the source, stated in \cite{LeTreust(EmpiricalCoordination)14}. 
\begin{eqnarray}
&&\max_{{\QQ}\in \Q_{\sf{sd}}} \bigg( I( W_1;Y  |V )  -   I(  U ; V  ,W_1  )  \bigg)>0.\label{eq:SCD}
\end{eqnarray}
\vspace{-1.3cm}

$\Q_{\sf{sd}}$ is the set of probability distributions ${\QQ}   \in  \Delta(\mc{U}   \times \mc{W}_1 \times \mc{X} \times \mc{Y} \times \mc{V}  )$ with auxiliary random variable $W_1$, that satisfy:
\begin{eqnarray*}
\PP_{u}(u)   \otimes  {\QQ}(x,v | u) \otimes  {\QQ}(w_1 | u,x,v) \otimes  \mc{T}(y | x )  .
\end{eqnarray*}
Equation \eqref{eq:SCDFeedbacks} corresponds to strictly causal decoding with feedback from the source, characterized in \cite{LarrousseLasaulceBloch(IT)14}.
\begin{eqnarray}
&&I(X  ;Y  |   U  ,V )   - I(U ;V ) >0.  \label{eq:SCDFeedbacks}
\end{eqnarray}
Equation \eqref{eq:SCDFeedbacks} can be deduced from equation \eqref{eq:SCD}, by replacing the auxiliary random variable $W_1$  by $X$ and the observation of the decoder $Y$ by the pair $(U,Y)$. 

\begin{figure}%[!ht]
\begin{center}
\psset{xunit=0.9cm,yunit=0.9cm}
\begin{pspicture}(0,-0.1)(8.5,0.9)
\pscircle(0,0.5){0.45}
\psframe(2,0)(3,1)
\pscircle(5,0.5){0.45}
\psframe(7,0)(8,1)
\psline[linewidth=1pt]{->}(0.5,0.5)(2,0.5)
\psline[linewidth=1pt]{->}(3,0.5)(4.5,0.5)
\psline[linewidth=1pt]{->}(5.5,0.5)(7,0.5)
\psline[linewidth=1pt]{->}(8,0.5)(9,0.5)
%\psline[linewidth=1pt]{->}(0,0)(0,-0.5)(5,-0.5)(5,0)
%\psline[linewidth=1pt]{->}(2.5,-0.5)(2.5,0)
%\psline[linewidth=1pt]{->}(0,1)(0,1.5)(7.5,1.5)(7.5,1)
\psline[linewidth=1pt]{->}(0,0)(0,-0.3)(7.5,-0.3)(7.5,0)
\rput[u](1,0.8){$U^n$}
%\rput[u](1,-0.2){$S^n$}
%\rput[u](1,1.8){$Z^{i}$}
\rput[u](3.75,0.8){$X^n$}
\rput[u](6.25,0.8){$Y^{i}$}
\rput[u](6.25,-0.0){$U^{i-1}$}
\rput[u](8.5,0.8){$V_i$}
\rput(0,0.5){$\PP_{\sf{u}}$}
\rput(2.5,0.5){$\C$}
\rput(5,0.5){$\mc{T}$}
\rput(7.5,0.5){$\D$}
\end{pspicture}
\caption{Non-causal encoding $f: \mc{U}^n  \rightarrow \mc{X}^n$ and causal decoding $g_i: \mc{Y}^i \times \mc{U}^{i-1} \rightarrow \mc{V}$ for all $i\in\{1,\ldots,n\}$ with feedback from the source.}\label{fig:CDf}
\end{center}
\end{figure}
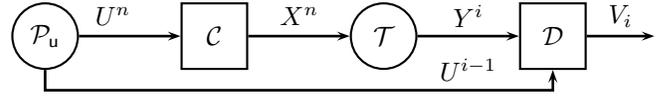

This analysis extends to causal decoding with feedback from the source, represented by Fig. \ref{fig:CDf} and characterized by  \eqref{eq:CDfeedbackSource}.
\begin{eqnarray}
&&\max_{{\QQ}\in \Q_{\sf{df}}} \bigg( I( X;Y  | U , W_3 )  -   I(  U ;W_3  )  \bigg)>0.\label{eq:CDfeedbackSource}
\end{eqnarray}
$\Q_{\sf{df}}$ is the set of probability distributions ${\QQ}   \in  \Delta(\mc{U}   \times \mc{W}_3 \times \mc{X} \times \mc{Y} \times \mc{V}  )$ with auxiliary random variable $W_3$, that satisfy:
\begin{eqnarray*}
\PP_{u}(u)   \otimes  {\QQ}(x,w_3 | u) \otimes  \mc{T}(y | x )  \otimes  {\QQ}(v | y,w_3) .
\end{eqnarray*}
The proof is in \cite{LeTreust(InternalRapportISITfeedbacks)15}. Theorems \ref{theo:feedbacks} and \ref{theo:Causalfeedbacks} also extend to two-sided state information by replacing $(U,S)$ by $(U,S,Y)$ in the results of \cite{LeTreust(ISIT-TwoSided)15}, for strictly causal and causal encoding.

  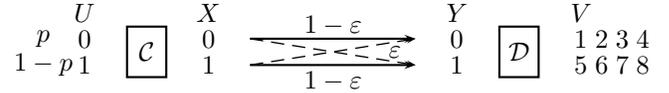
\begin{figure}[ht]
\begin{center}
\psset{xunit=0.55cm,yunit=0.35cm}
\begin{pspicture}(-1,0.5)(15,2.2)
\rput(0,1.5){$p$}
\rput(0,0.5){$1-p$}
\rput(1,2.5){$U$}
\rput(1,1.5){$0$}
\rput(1,0.5){$1$}
\psframe(2,0)(3,2)
\rput(2.5,1){$\C$}
\rput(4,2.5){$X$}
\rput(4,1.5){$0$}
\rput(4,0.5){$1$}
\psline{->}(5,1.5)(9,1.5)
\psline{->}(5,0.5)(9,0.5)
\psline[linewidth=0.5pt, linestyle = dashed]{->}(5,1.5)(9,0.5)
\psline[linewidth=0.5pt, linestyle = dashed]{->}(5,0.5)(9,1.5)
\rput(7,2){$1 - \varepsilon$}
\rput(7,-0.1){$1 - \varepsilon$}
\rput(8.5,1){$\varepsilon$}
\rput(10,2.5){$Y$}
\rput(10,1.5){$0$}
\rput(10,0.5){$1$}
\psframe(11,0)(12,2)
\rput(11.5,1){$\D$}
\rput(13,2.5){$V$}
\rput(13,1.5){$1$}
\rput(13,0.5){$5$}
\rput(13.5,1.5){$2$}
\rput(13.5,0.5){$6$}
\rput(14,1.5){$3$}
\rput(14,0.5){$7$}
\rput(14.5,1.5){$4$}
\rput(14.5,0.5){$8$}
%\rput(13,1.5){$1$}
%\rput(13,0.5){$2$}
%\rput(13.5,1.5){$3$}
%\rput(13.5,0.5){$4$}
%\rput(14,1.5){$5$}
%\rput(14,0.5){$6$}
%\rput(14.5,1.5){$7$}
%\rput(14.5,0.5){$8$}
\end{pspicture}
\caption{Binary information source and binary symmetric channel with parameters $p=1/2$ and $\varepsilon \in [0,0.5]$}\label{fig:JointSourceChannelProblem2x2b}
\end{center}
\end{figure}

\vspace{-0.5cm}

\section{Example: binary source and channel}\label{sec:NumericalExample}
We consider a binary information source and a binary symmetric
\begin{figure}[ht]
\begin{center}
\psset{xunit=0.6cm,yunit=0.5cm}
\begin{pspicture}(-1.5,-0.3)(6,8)
\psdots(0,0)(0,1)(0,2)(0,3)(0,4)(0,5)(0,6)(0,7)
\rput(-1.5,8){$U,X,Y$}
\rput(-1.5,7){$(0,0,0)$}
\rput(-1.5,6){$(1,0,0)$}
\rput(-1.5,5){$(0,1,0)$}
\rput(-1.5,4){$(1,1,0)$}
\rput(-1.5,3){$(0,0,1)$}
\rput(-1.5,2){$(1,0,1)$}
\rput(-1.5,1){$(0,1,1)$}
\rput(-1.5,0){$(1,1,1)$}
\psline{->}(0.5,0)(4.5,0)
\psline{->}(0.5,1)(4.5,1)
\psline{->}(0.5,2)(4.5,2)
\psline{->}(0.5,3)(4.5,3)
\psline{->}(0.5,4)(4.5,4)
\psline{->}(0.5,5)(4.5,5)
\psline{->}(0.5,6)(4.5,6)
\psline{->}(0.5,7)(4.5,7)
\psdots(5.3,0)(5.3,1)(5.3,2)(5.3,3)(5.3,4)(5.3,5)(5.3,6)(5.3,7)
\rput(6,8){$V$}
\rput(6,7){$1$}
\rput(6,6){$2$}
\rput(6,5){$3$}
\rput(6,4){$4$}
\rput(6,3){$5$}
\rput(6,2){$6$}
\rput(6,1){$7$}
\rput(6,0){$8$}
\rput(2.5,7.4){$1 - \alpha$}
\rput(4.5,6.5){$ \alpha / 7$}
\rput(4.5,5.5){$ \alpha / 7$}
\rput(4.5,4.5){$ \alpha / 7$}
\rput(4.5,3.5){$ \alpha / 7$}
\rput(4.5,2.5){$ \alpha / 7$}
\rput(4.5,1.5){$ \alpha / 7$}
\rput(4.5,0.5){$ \alpha / 7$}
\psline[linewidth=0.5pt, linestyle = dashed]{->}(0.5,7)(4.5,6)
\psline[linewidth=0.5pt, linestyle = dashed]{->}(0.5,7)(4.5,5)
\psline[linewidth=0.5pt, linestyle = dashed]{->}(0.5,7)(4.5,4)
\psline[linewidth=0.5pt, linestyle = dashed]{->}(0.5,7)(4.5,3)
\psline[linewidth=0.5pt, linestyle = dashed]{->}(0.5,7)(4.5,2)
\psline[linewidth=0.5pt, linestyle = dashed]{->}(0.5,7)(4.5,1)
\psline[linewidth=0.5pt, linestyle = dashed]{->}(0.5,7)(4.5,0)
\end{pspicture}
\caption{Conditional probability distribution $\QQ_{\sf{v|uxy}}$ depending on parameter $\alpha \in [0,7/8]$ where $\QQ\big(V=1\big|(U,X,Y)=(0,0,0)\big) = 1 - \alpha$ and $\QQ\big(V=2\big|(U,X,Y)=(0,0,0)\big) =  \alpha / 7$. For $\alpha = 7/8$, the probability distribution is uniform over the set $\mc{V} = \{1,\ldots,8\}$ and independent of the triple $(U,X,Y)$. For $\alpha = 0$, the output $V$ corresponds exactly to the triple $(U,X,Y)$.}\label{fig:ConditionalProba}
\end{center}
\end{figure}
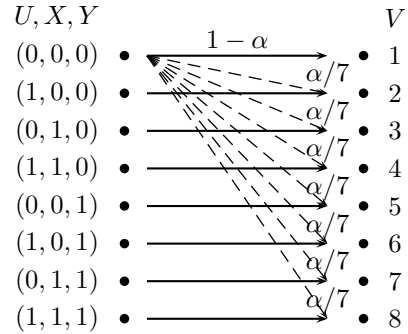
channel represented by Fig. \ref{fig:JointSourceChannelProblem2x2b}. The set of symbols are 
given by $\mc{U} = \mc{X} = \mc{Y} = \{0,1\}$ and $\mc{V} = \{1,2,3,4,5,6,7,8\}$. We assume the parameter $p\in [0,1]$ of the information source is equal to 1/2. The probability distribution of channel input is uniform $\QQ(X=0) = \QQ(X = 1) = 1/2$. The transition probability of the channel depends on a noise parameter $\varepsilon \in [0,0.5]$.  Since the input distribution is uniform and the channel is symmetric, the output probability distribution is also uniform $\QQ(Y=0) = \QQ(Y = 1) = 1/2$. We investigate a class of achievable conditional probability distributions $\QQ_{\sf{v|uxy}}$ described by Fig. \ref{fig:ConditionalProba}. 

We consider strictly causal encoding with feedback. The information constraint \eqref{eq:Feedbacks1} of Theorem  \ref{theo:feedbacks} writes:
\footnotesize
\begin{eqnarray*}
&&I( X;Y  )  -  I(  U;V |X,Y ) \\
&=&H(Y) - H(Y|X) - H(V|X,Y) + H(V |U,X,Y) \\
&=&1 - H_b(\varepsilon) - H_b\bigg( \frac{6\alpha}{7}  \bigg) - 1 -  \frac{6\alpha}{7} \cdot \log_2 3  + H_b(\alpha) +  \alpha \cdot \log_2 7\\
&=&  H_b(\alpha) - H_b(\varepsilon) - H_b\bigg( \frac{6\alpha}{7}  \bigg)   +   \alpha \cdot \bigg( \log_2 7 -  \frac{6}{7} \cdot \log_2 3 \bigg) .
\end{eqnarray*}

\vspace{-0.7cm}

\begin{figure}[ht!]
\centering
\includegraphics[width=0.4\textwidth]{%/Users/maelletreust/Documents/MATLAB/
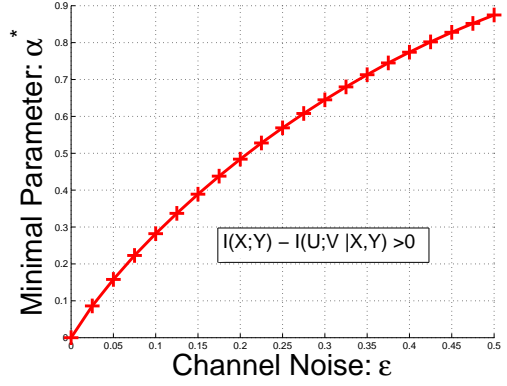}
\caption{Minimal value of parameter $\alpha^{\star}  \in [0,0.875]$ for the information constraint $I(X;Y) - I(U;V|X,Y)>0$ to be positive, depending on the noise of the channel $\varepsilon \in [0,0.5]$. It corresponds to the higher level of coordination between the random variable $V$ and the triple $(U,X,Y)$.}\label{fig:CoordinationFeedbacksB_2015_01_14}
\end{figure}

\normalsize
\begin{figure}[ht!]
\centering
\includegraphics[width=0.45\textwidth]{%/Users/maelletreust/Documents/MATLAB/
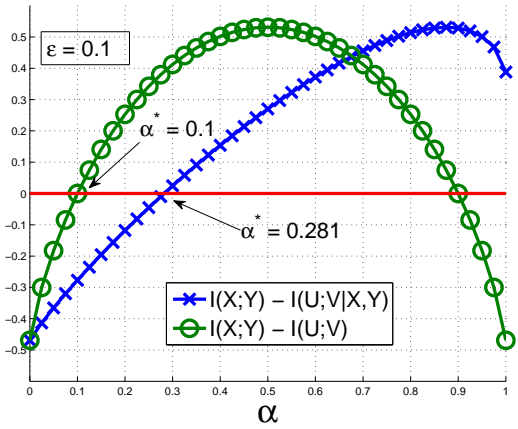}
\caption{Comparison between the information constraint for empirical coordination with feedback $I(X;Y) - I(U;V|X,Y)$ and the information constraint $I(X;Y) - I(U;V)$ for lossy transmission.}\label{fig:CoordinationFeedbacks2015_01_14}
\end{figure}
In Fig. \ref{fig:CoordinationFeedbacks2015_01_14}, we compare the information constraint for empirical coordination with feedback \eqref{eq:Feedbacks1} and information constraint for lossy transmission without coordination \eqref{eq:IClossy}, where $\alpha$ is the distortion parameter of conditional distribution $\QQ_{\sf{v|u}}$:
\begin{eqnarray}
I( X;Y  )  -  I(  U;V  )&=& 1 - H_b(\varepsilon) - 1 + H_b (\alpha) \label{eq:IClossy}\\
&=& H_b (\alpha) - H_b(\varepsilon) .
\end{eqnarray}
%This correspond to the lossy transmission of the information source $U$ over the noisy channel $\mc{T}_{\sf{y|x}}$. 
%The minimal distortion $\alpha$ of the source correspond to the error probability of the channel $\varepsilon = \alpha$. Fig. \ref{fig:CoordinationFeedbacks2015_01_14} illustrate that 
The minimal coordination parameter $\alpha^{\star} \simeq 0.281 >0.1$ is much larger for empirical coordination than for lossy compression. This restriction comes from the additional correlation requirement between the decoder output $V$ and the random variables $(X,Y)$ of the channel. Fig. \ref{fig:CoordinationFeedbacksB_2015_01_14} provides the minimal value of parameter $\alpha^{\star} \in [0,0.875]$ for empirical coordination, depending on the level of noise of the channel $\varepsilon \in [0,0.5]$.

%%%%%%%%%%%%%%%%%%%%%%%%%%%%%%%%%%%%%%
%%%%%%%%%%%%%%%%%%%%%%%%%%%%%%%%%%%%%%

\section{Conclusion}\label{sec:Conclusion}

We investigate the relationship between coordination and feedback by considering a point-to-point scenario with strictly causal and causal encoder. For both cases, we characterize the optimal solutions and we show that feedback simplifies the information constraints by reducing the number of auxiliary random variables. For empirical coordination with strictly causal encoding and feedback, the information constraint does not involve auxiliary random variable anymore.

%%%%%%%%%%%%%%%%%%%%%%%%%%%%%%%%%%%%%%%
%%%%%%%%%%%%%%%%%%%%%%%%%%%%%%%%%%%%%%%

\appendix\label{sec:App}

The full versions of the proofs are stated in \cite{LeTreust(InternalRapportISITfeedbacks)15}.

%%%%%%%%%%%%%%%%%%%%%%%%%%%%%%%%%%%%%%%
%%%%%%%%%%%%%%%%%%%%%%%%%%%%%%%%%%%%%%%

\subsection{Sketch of proof of Theorem \ref{theo:feedbacks}}\label{sec:ProofFeedbacks2}

Achievability proof can be obtained from the proof of Theorem \ref{theo:Causalfeedbacks} stated in Appendix \ref{sec:ProofFeedbacks3}, by replacing the auxiliary random variable $W$ by $X$.

For the converse proof, we consider code $c(n) \in \mc{C}$ with small error probability $\PP_{\sf{e}}(c)$.
\scriptsize
\begin{eqnarray}
0&=&   I(U^n ; Y^n)  -  I(U^n ; Y^n,V^n)    \label{eq:ConvFeedbacks2} \\
&=& I(U^n ; Y^n)  -  \sum_{i=1}^n I(U_i ; Y^n,V^n ,  U^{i-1})    \label{eq:ConvFeedbacks5} \\
&\leq& \sum_{i=1}^n \bigg( I(Y_i ;U^n, X_i | Y^{i-1})  -  I(U_i ; Y^n,V^n ,  U^{i-1}, X_i)   \bigg)\label{eq:ConvFeedbacks6} \\
&\leq&
%\sum_{i=1}^n \bigg(H( Y_i | Y^{i-1}) - H(Y_i | Y^{i-1} , U^n )  -   I(U_i ; Y_i,V_i ,  X_i)  \bigg) \label{eq:ConvFeedbacks8} \\
%&=&
\sum_{i=1}^n \bigg(H( Y_i) - H(Y_i | X_i )  -   I(U_i ; Y_i,V_i ,  X_i)  \bigg) \label{eq:ConvFeedbacks9} \\
&\leq&  \sum_{i=1}^n \bigg(H( Y_i)+ H(U_i | Y_i,V_i ,  X_i)   \bigg) - n \bigg(  H( Y | X ) + H(U) \bigg) \label{eq:ConvFeedbacks10} \\
&\leq&  n \bigg(  I(X; Y) - I(U ; V | X,Y)  \bigg) \label{eq:ConvFeedbacks11} .
\end{eqnarray}
\normalsize
Equation \eqref{eq:ConvFeedbacks2} comes from the non-causal decoding that induces the Markov chain: $U^n -\!\!\!\!\minuso\!\!\!\!-    Y^n -\!\!\!\!\minuso\!\!\!\!-  V^n $.\\
Equation \eqref{eq:ConvFeedbacks5} comes from the i.i.d. properties of the information source $U$ that implies: $I(U_i  ; U^{i-1} )=0$.\\
Equation \eqref{eq:ConvFeedbacks6} comes from the channel feedback and the strictly causal encoding function: $X_i = f_i(U^{i-1} ,  Y^{i-1})$.\\
%Equation \eqref{eq:ConvFeedbacks8} comes from the properties of the mutual information.\\
%Equation \eqref{eq:ConvFeedbacks9} is due to the memoryless prop. of the channel.\\
%strictly causal encoding function $X_i = f_i(U^{i-1} ,  Y^{i-1})$ and the memoryless property of the channel that imply: $H(Y_i | Y^{i-1} , U^n )= H(Y_i | Y^{i-1} , U^n,X_i ) = H(Y_i | X_i )$.\\
Equations \eqref{eq:ConvFeedbacks9} and \eqref{eq:ConvFeedbacks10} are due to the properties of i.i.d. information source and of memoryless channel.\\
Equation \eqref{eq:ConvFeedbacks11} comes from the concavity of the entropy function and from the hypothesis of small error probability $\PP_{\sf{e}}(c)$.

\subsection{Sketch of achievability proof of Theorem \ref{theo:Causalfeedbacks}}\label{sec:ProofFeedbacks3}

Consider $\QQ \in \Q$ that achieves the maximum in equation \eqref{eq:FeedbacksCausal1}. There exists a $\delta>0$ and a rate $ \textsf{R} >0$ such that:
%
%\begin{eqnarray}
%I(W ; Y  )   -  I( U;V  | W , Y )   &=&  I(W ; Y  ) +  I(V ; Y | W ) -  I(V ; Y | W )  -  I( U;V  | W , Y ) \\
% &=&   I(W , V ; Y  )   -  I( U, Y ;V  | W ) >0 . \label{eq:Feedbacks3}
%\end{eqnarray}
%There exists a $\delta>0$ and a rate $ \textsf{R} >0$ such that:
\small
\begin{eqnarray}
\textsf{R}  & \geq&   I( U, Y ;V  | W ) + \delta   , \label{eq:AchCausalFeed1}\\
\textsf{R}   & \leq&   I(W  ; Y  ) +  I( V ; Y | W ) -  \delta =   I(W , V ; Y  )  -  \delta . \label{eq:AchCausalFeed2}
\end{eqnarray}
\normalsize
We define a block-Markov random code $c\in \mc{C}(n)$ over $B\in \N$ blocks of length $n\in \N$. 
%The total length of the code is denoted by $N = n\cdot B\in \N$  and $\sf{R}$ denotes the rate of the code.\\
\begin{itemize}
\item[$\bullet$] \textit{Random codebook.} We generate $| \mc{M}  |= 2^{n   \sf{R}  } $ sequences $W^n(m)$  drawn from $\QQ_{\sf{w}}^{\otimes n} $ with index  $m\in \mc{M} $. For each index  $m\in \mc{M} $, we generate the same number $| \mc{M}  |= 2^{n   \sf{R} } $ of sequences $ V^n(m,\hat{m})$ with index  $\hat{m} \in \mc{M} $, drawn from $\QQ_{\sf{v|w}}^{\otimes n} $ depending on  $W^n(m)$.  
\item[$\bullet$] \textit{Encoding function.} It recalls $m_{b-1} $ and finds
$ m_b\in \mc{M}$ s.t. sequences $(U^n_{b-1},Y^n_{b-1}, W^n(m_{b-1}) ,V^n(m_{b-1},m_b) )\in A_{\varepsilon}^{{\star}{n}}(\QQ)$ are jointly typical in block $b-1$. It deduces $W^n(m_b) $ for block $b$ and sends $X_b^n$ drawn from $\QQ_{\sf{x|uw}}^{\otimes n}$  depending on $(U^n_b, W^n(m_b))$. 
\item[$\bullet$] \textit{Decoding function.} It recalls $m_{b-1}$ and finds
$ m_b\in \mc{M}$ s.t. sequences $(Y^n_b , W^n(m_b) ) \in A_{\varepsilon}^{{\star}{n}}( \QQ)$  and  $(Y^n_{b-1} ,W^n(m_{b-1}), V^n(m_{b-1} , m_b) ) \in A_{\varepsilon}^{{\star}{n}}( \QQ)$ are jointly typical. It returns $V^n(m_{b-1} , m_b) $ over block $b-1$.
\item[$\bullet$] \textit{First block at the encoder.} An arbitrary index $m_1 \in \mc{M} $ of $W^n(m_1) \in \mc{W}^n$ is given to encoder and decoder. Encoder sends $X_{b_1}^n$ drawn from $\QQ_{\sf{x|uw}}^{\otimes n}$  depending on  $( U^n_{b_1}, W^n(m_{1}))$. At the beginning of the second block $b_2$, encoder %recalls $(U^n_{b_1}, Y^n_{b_1}, W^n(m_{1}))$ and 
finds index $m_2$ such that  $(U^n_{b_1},Y^n_{b_1}, W^n(m_{1}) ,V^n(m_{1},m_2) )\in A_{\varepsilon}^{{\star}{n}}(\QQ)$. It sends $X_{b_2}^n$ drawn from $\QQ_{\sf{x|wu}}^{\otimes n}$  depending on $( U^n_{b_2}  , W^n(m_{2}))$.
\item[$\bullet$] \textit{First block at the decoder.}  At the end of second block $b_2$, the decoder finds the index $m_2$ such that $(Y^n_{b_2} , W^n(m_2) ) \in A_{\varepsilon}^{{\star}{n}}( \QQ)$  and  $(Y^n_{b_1} , W^n(m_{1}), V^n(m_{1} , m_2) ) \in A_{\varepsilon}^{{\star}{n}}( \QQ)$. Over the first bloc, decoder $\D$ returns $V^n(m_{1} , m_2)\in \mc{V}^n$. Sequences $(U^n_{b_1} , W^n(m_{1}), X^n_{b_1}, Y^n_{b_1} , V^n(m_{1} , m_2)) \in A_{\varepsilon}^{{\star}{n}}( \QQ)$ are jointly typical over the first block $b_1$.
\item[$\bullet$] \textit{Last bloc.}  Sequences are not jointly typical.
\end{itemize}
Equations \eqref{eq:AchCausalFeed1}, \eqref{eq:AchCausalFeed2} imply for all $n\geq\bar{n}$, for a large number of blocks $B \in \N$, the sequences are jointly typical with large probability.
\tiny
\begin{eqnarray*}
&&\E_c\bigg[ \PP\bigg( U^n \notin A_{\varepsilon}^{{\star}{n}}(\QQ) \bigg)\bigg]  \leq \varepsilon,\\
&&\E_c\bigg[ \PP\bigg( \forall  m \in \mc{M}  ,\; 
(U^n_{b-1},Y^n_{b-1}, W^n(m_{b-1}) ,V^n(m_{b-1},m) ) \notin A_{\varepsilon}^{{\star}{n}}(\QQ) \bigg)\bigg]  \leq \varepsilon,\\
&&\E_c\bigg[ \PP\bigg(  \exists m'\neq  m  ,\text{ s.t. } 
\Big\{ (Y^n_b   ,W^n(m') ) \in A_{\varepsilon}^{{\star}{n}}( \QQ)\Big\} \cap \nonumber \\
&&  \qquad \qquad\qquad\Big\{(Y^n_{b-1} ,W^n(m_{b-1}), V^n(m_{b-1} , m') ) \in A_{\varepsilon}^{{\star}{n}}( \QQ) \Big\} \bigg)\bigg]   \leq \varepsilon.\label{eq:achievDoublecausal}
\end{eqnarray*}
\normalsize

\subsection{Sketch of Converse Proof of Theorem \ref{theo:Causalfeedbacks}}\label{sec:ProofFeedbacks4}

Consider code $c(n) \in \mc{C}$ with small error probability $\PP_{\sf{e}}(c)$.
\tiny
\begin{eqnarray}
0&\leq&  
%\sum_{i=1}^n I(   U^{i-1} , Y^{i-1}   ;  Y_i  | Y^n_{i+1}   )  -   \sum_{i=1}^n I(Y^n_{i+1}  ;  U_i  ,Y_i |  U^{i-1} , Y^{i-1} )    \label{eq:ConvCausalEncodingF1} \\
%&\leq&
  \sum_{i=1}^n I(   U^{i-1}   , Y^{i-1} , Y^n_{i+1}  ;  Y_i  )  -   \sum_{i=1}^n I(Y^n_{i+1}    ;  U_i  ,Y_i |  U^{i-1} , Y^{i-1} )  \label{eq:ConvCausalEncodingF2} \\
&=&  \sum_{i=1}^n I(   U^{i-1}   , Y^{i-1}  ;  Y_i  )  -   \sum_{i=1}^n I(Y^n_{i+1}    ;  U_i   |  U^{i-1} , Y^{i-1} , Y_i )  \label{eq:ConvCausalEncodingF3} \\
&=&  \sum_{i=1}^n I(   U^{i-1}   , Y^{i-1}  ;  Y_i  )  -   \sum_{i=1}^n I(Y^n_{i+1}   ,V_i ;  U_i   |  U^{i-1} , Y^{i-1} , Y_i )  \label{eq:ConvCausalEncodingF4} \\
&\leq&  \sum_{i=1}^n I(   U^{i-1}   , Y^{i-1}  ;  Y_i  )  -   \sum_{i=1}^n I(V_i ;  U_i   |  U^{i-1} , Y^{i-1} , Y_i )  \label{eq:ConvCausalEncodingF5} \\
&=&  \sum_{i=1}^n I(  W_{i}     ;  Y_i   )   -   \sum_{i=1}^n I( V_i  ;  U_i  |  W_i , Y_i ) .    \label{eq:ConvCausalEncodingF6} \\
&\leq&  n \cdot  \max_{{\QQ}\in \Q}  \bigg( I(  W     ;  Y   )   -   I( V  ;  U   |  W , Y  )   \bigg)  \label{eq:ConvCausalEncodingF10} .
\end{eqnarray}
\normalsize
%Equation \eqref{eq:ConvCausalEncodingF2} is due to Csisz\'{a}r Sum Identity and the properties of the mutual information.\\
Eq. \eqref{eq:ConvCausalEncodingF2}, \eqref{eq:ConvCausalEncodingF3} are due to Csisz\'{a}r Sum Identity, prop. of MI.\\
Eq. \eqref{eq:ConvCausalEncodingF4} is due to the non-causal decoding function $V^n = g(Y^n)$, that implies: $I(V_i ;  U_i   |  U^{i-1} , Y^{i-1} , Y_i , Y^n_{i+1} ) = 0$.\\
Eq.  \eqref{eq:ConvCausalEncodingF5} is due to the properties of the mutual information.\\
Eq. \eqref{eq:ConvCausalEncodingF6} is due to the introduction of auxiliary random variables $ W_{i} =  (U^{i-1} , Y^{i-1})$ satisfying properties of set $\Q$.\\
Eq. \eqref{eq:ConvCausalEncodingF10} comes from taking the maximum over the set $\Q$.
%. For all $i \in \{1,\ldots,n\}$, the auxiliary random variable $ W_i $ satisfies the properties corresponding to the  set of probability distribution $\Q$:
\begin{eqnarray}
&& U_i \text{ is independent of } W_{i}, \label{eq:ConvCEMarkovF1} \\
&& Y_i -\!\!\!\!\minuso\!\!\!\!-   X_i  -\!\!\!\!\minuso\!\!\!\!-  (U_i , W_i) , \label{eq:ConvCEMarkovF3}  \\
&& V_i -\!\!\!\!\minuso\!\!\!\!-   (U_i, Y_i,  W_i) -\!\!\!\!\minuso\!\!\!\!- X_i. \label{eq:ConvCEMarkovF5} 
\end{eqnarray}
$\bullet$ Eq. \eqref{eq:ConvCEMarkovF1} is due to the i.i.d. property of the source that implies $U_i$ is independent of $U^{i-1}$. The causal encoding with feedback $X_i = f_i(U^i, Y^{i-1})$ and the memoryless property of the channel implies that $Y^{i-1}$ is independent of $U_i$. \\
$\bullet$ Eq. \eqref{eq:ConvCEMarkovF3} comes from the memoryless property of the channel and the fact that $Y_i$ is not included in $W_i$.\\
$\bullet$ Eq. \eqref{eq:ConvCEMarkovF5} comes from the causal encoding with feedback function that implies that $X_i$  is a deterministic function of $(U_i,U^{i-1} ,Y^{i-1} )$ which is included in $(U_i ,Y_i, W_i )$.\\

  %%%%%%%%%%%%%%%%%%%%%%%%%%%%%%%%%%%%%%%%%%
 %%%%%%%%%%%%%%%%%%%%%%%%%%%%%%%%%%%%%%%%%%
 %%%%%%%%%%%%%%%%%%%%%%%%%%%%%%%%%%%%%%%%%%

 %%%%%%%%%%%%%%%%%%%%%%%%%%%%%%%%%%%%%%%%%%%%%%%%%%%%%%
%%%%%%%%%%%%%%%%%%%%%%%%%%%%%%%%%%%%%%%%%%%%%%%%%%%%%%
%%%%%%%%%%%%%%%%%%%%%%%%%%%%%%%%%%%%%%%%%%%%%%%%%%%%%%
%%%%%%%%%%%%%%%%%%%%%%%%%%%%%%%%%%%%%%%%%%%%%%%%%%%%%%

%
%%%
%\bibliographystyle{ieeetr}
%\bibliography{/Users/maelletreust/Documents/Redaction/BiblioMael}
%
\vspace{-0.5cm}

%%%%%%%%

%%%%%%%%

\end{document}